\documentstyle[prc,aps]{revtex}
\title{Charge-Asymmetry of the Nucleon-Nucleon Interaction}
\author{G. Q. Li\thanks{Present Address: Department of Physics,
SUNY, Stony Brook, NY 11794} and
R. Machleidt\thanks{Electronic address: machleid@uidaho.edu}}
\address{\it Department of Physics, University of Idaho,
Moscow, ID 83844, U.S.A.}
\date{\today}

\begin{document}

\maketitle

\begin{abstract}
Based upon the Bonn meson-exchange model 
for the nucleon-nucleon ($NN$) interaction, 
we study systematically the charge-symmetry-breaking (CSB)
of the $NN$ interaction due to nucleon mass splitting.
Particular attention is payed to CSB generated by
the $2\pi$-exchange contribution to the $NN$ interaction,
$\pi\rho$ diagrams, and other multi-meson-exchanges.
We calculate the CSB differences in the $^1S_0$ 
effective range parameters
as well as phase shift differences in $S$, $P$ and higher partial
waves up to 300 MeV lab.\ energy. 
We find a total CSB difference in the singlet scattering length of
1.6 fm which explains the empirical value accurately.
The corresponding CSB phase-shift differences
are appreciable at low energy in the $^1S_0$ state.
In the other partial waves, the CSB splitting of the phase shifts
is small and increases with energy, with typical values in the order of
0.1 deg at 300 MeV in $P$ and $D$ waves.
\end{abstract}
\pacs{PACS numbers: 24.80.+y, 11.30.Hv, 13.75.Cs, 21.30.Cb}

\twocolumn

\section{Introduction}
Charge-symmetry is the equality of proton-proton ($pp$)
and neutron-neutron ($nn$) forces---after electromagnetic effects
are removed.
This symmetry, which is slightly broken,
has long been a subject of
research in nuclear physics 
(for reviews see, e.~g., Refs.~\cite{SAT89,MNS90,MO95,Mac89}). 
Traditionally, the empirical information on the 
charge-asymmetry of the nuclear force comes mainly
from few-body systems. The nucleon-nucleon ($NN$)
 scattering length in the $^1S_0$ state plays
a special role. As there exists an almost bound state in that partial wave, the
(negative) scattering length is extremely sensitive to small differences in the
strength of the force. 
The $pp$ effective range parameters (scattering length, $a$,
and effective range, $r$) are obtained with very high 
precision from  low-energy $pp$ cross section data.
However, since we are interested here in the strong force,
electromagnetic effects have to be removed, which introduces model
dependence. Using several realistic $NN$ potential models, the pure
 strong-interaction
 $pp$ effective range parameters are determined to be~\cite{MNS90}
\begin{eqnarray}
a_{pp}^N& =& -17.3 \pm 0.4 \mbox{ fm} , 
\\
r_{pp}^N& =& 2.85 \pm 0.04 \mbox{ fm} ,
\end{eqnarray}
where the errors state the uncertainty due to model-dependence.

Since $nn$ scattering experiments are not yet feasible,
the $nn$ effective range parameters are not
measured directly; they are extracted from few-body reactions,
mainly $D(n,nn)p$ and $D(\pi^-,\gamma)2n$. 
Recent measurements of these reactions and their analysis
have resulted in the following recommended values~\cite{SAT89,MNS90}
\begin{eqnarray}
a_{nn}^N& =& -18.8\pm 0.3 \mbox{ fm} , 
\\
r_{nn}^N& =& 2.75\pm 0.11 \mbox{ fm} .
\end{eqnarray}
It is thus evident that in the $^1S_0$ state, the 
$nn$ strong interaction is slightly
more attractive than the $pp$ one. 
From the above semi-empirical values, we see that
charge-symmetry is broken by the following amounts
\begin{eqnarray}
\Delta a_{CSB}& \equiv& a_{pp}^N-a_{nn}^N = 1.5\pm 0.5 \mbox{ fm}, \\
\Delta r_{CSB}& \equiv& r_{pp}^N-r_{nn}^N = 0.10\pm 0.12 \mbox{ fm}. 
\end{eqnarray}

Information about charge-symmetry-breaking (CSB)
 can also be inferred from binding
energy differences of so-called mirror nuclei. The most studied case
is the $^3$He--$^3$H mirror pair. Experimentally it was found  that
$^3$H is more deeply bound than $^3$He by 764 keV. 
Model-independent calculations of the Coulomb energy difference and other
subtle electromagnetic effects
yield a binding energy difference
of about 683$\pm$29 keV \cite{BCS78}. 
It has been shown that the remaining discrepancy can be explained by
a charge symmetry breaking nuclear force that is consistent with the 
empirical asymmetry
in the singlet scattering length \cite{Bra88}. 

According to our current understanding,
CSB is due to a mass difference between the up and down quark and
electromagnetic interactions. On the hadronic
level, this has various consequences: mixing of mesons
of different isospin but same spin and parity; 
mass differences between hadrons of the same isospin multiplet.

The difference between the masses of  neutron and proton
represents the most basic cause for CSB. Therefore, it is 
important to have a very thorough accounting of this effect.
This is the subject of the present paper.

The $n-p$ mass difference, 
which is well known to be 1.2933 MeV~\cite{PDG92}, affects the
kinetic energy of the nucleons.
Besides this, it has also an impact on all meson-exchange
diagrams that contribute to the nuclear force.

In Sect.~II, we will briefly outline the formalism of the Bonn
model for the $NN$ interaction that this study is based upon.
In Sect.~III, we will go---step by step---through the various 
meson-exchange contributions to the nuclear force and calculate
for each step the CSB effect
due to nucleon mass splitting.
In particular, we will present the effect on the singlet effective
range parameters and on phase shifts of $NN$
scattering up to 300 MeV laboratory energy and up to
orbital angular momentum $L=2$.
Section IV concludes the paper.

\section{Sketch of model}

We base our investigation on the comprehensive Bonn full model for the $NN$ 
interaction.
This model has been described in length in the 
literature~\cite{Mac89,MHE87,ML94}.
Therefore, we will summarize here only those facts which are important for
the issue under consideration.

The Bonn model uses an effective, field-theoretic approach,
in which the interaction between two nucleons is created solely from
the exchange of mesons; 
namely, 
$\pi$, $\rho(770)$, $\omega(782)$, $a_0/\delta(980)$,
and $\sigma'(550)$.
Besides the nucleon, also the $\Delta (1232)$ isobar is taken into account.
In its original version~\cite{MHE87}, the Bonn model
used averages for baryon and meson masses and, thus, was charge-independent;
it was fitted to the neutron-proton data.
In this paper, these subtleties will be treated accurately.

The interaction Lagrangians involving pions are
\begin{eqnarray}
{\cal L}_{\pi NN} & = &  \frac{f_{\pi NN}}{m_{\pi^\pm}}
                           \bar{\psi}  \gamma_\mu \gamma_5 
                           \mbox{\boldmath $\tau$} \psi
                           \cdot \partial^\mu
                           \mbox{\boldmath $\varphi$}_\pi \; ,
\\
{\cal L}_{\pi N\Delta} & = & \frac{f_{\pi N\Delta}}{m_{\pi^\pm}}
                           \bar{\psi} 
                           \mbox{\boldmath $T$} \psi_\mu
                           \cdot \partial^\mu
                           \mbox{\boldmath $\varphi$}_\pi 
                            + \mbox{ H.c.} \; ,
\end{eqnarray}
with $\psi$ the nucleon, $\psi_\mu$ the $\Delta$ (Rarita-Schwinger
spinor), and 
$\mbox{\boldmath $\varphi$}_\pi$ the pion fields.
{\boldmath $\tau$} are the usual Pauli matrices describing
isospin 1/2 and 
{\boldmath $T$}
is the isospin transition operator.
H.c. denotes the Hermitean conjugate.

The above Lagrangians are devided by $m_{\pi^\pm}$ to make
the coupling constants $f$ dimensionless.
Following established conventions~\cite{Dum83},
we always use $m_{\pi^\pm}$ as scaling mass.
It may be tempting to use $m_{\pi^0}$ for $\pi^0$ coupling.
Notice, however, that the scaling mass could be anything.
Therefore, it is reasonable to keep the scaling mass constant
within SU(3) multiplets~\cite{Dum83}. This avoids the creation of
unmotivated charge-dependence. 

It is important to stress that---as evidenced by the
above $\pi NN$ Langrangian---we use the pseudovector (pv)
or gradient coupling for the pion. Alternatively, on can
also use the pseudoscalar (ps) coupling,
\begin{equation}
{\cal L}_{\pi NN}^{(ps)}  =   g_{\pi NN}
                           \bar{\psi}  i \gamma_5 
                           \mbox{\boldmath $\tau$} \psi
                           \cdot 
                           \mbox{\boldmath $\varphi$}_\pi \; .
\end{equation}

For an on-shell process, the two couplings yield the same if
the coupling constants are related by,
\begin{equation}
g_{\pi NN} = \left( \frac{M_1+M_2}{m_{\pi^\pm}} \right)
 f_{\pi NN} \; ,
\end{equation}
with $M_1$ and $M_2$ the masses of the two nucleons involved.
This relationship 
is charge-dependent 
due to the two nucleon masses. 
As a consequence, CSB effects will come
out (noticably!) different depending on if the ps or the pv coupling
is used.
Non-linear realizations of chiral symmetry, which are currently
fashionable, prefer the pv coupling over the ps coupling. 
Following this trend, we use the pv coupling.

The couplings of $\rho$-mesons to nucleons and $\Delta$-isobars are described
by the Lagrangians
\begin{eqnarray}
{\cal L}_{\rho NN}& =& g_{\rho NN}
                           \bar{\psi}  \gamma_\mu  
                           \mbox{\boldmath $\tau$} \psi
                           \cdot 
                           \mbox{\boldmath $\varphi$}_{\rho}^\mu
\nonumber
\\
           &  &   + \frac{f_{\rho NN}}{4M_p}
                           \bar{\psi}  \sigma_{\mu\nu}  
                           \mbox{\boldmath $\tau$} \psi
                           \cdot 
                           (\partial^\mu
                           \mbox{\boldmath $\varphi$}_{\rho}^\nu
                           -\partial^\nu
                           \mbox{\boldmath $\varphi$}_{\rho}^\mu)
 \; ,
\\
{\cal L}_{\rho N\Delta} & = & i \frac{f_{\rho N\Delta}}{m_{\rho^\pm}}
                           \bar{\psi} \gamma_5 \gamma_\mu
                           \mbox{\boldmath $T$} \psi_\nu
                           \cdot 
                           (\partial^\mu
                           \mbox{\boldmath $\varphi$}_{\rho}^\nu
                           -\partial^\nu
                           \mbox{\boldmath $\varphi$}_{\rho}^\mu)
                            + \mbox{ H.c.} \; .
\end{eqnarray}
We have to draw attention to the fact that---no matter
to which nucleon the $\rho$ couples---in the second part
of the $\rho NN$ Langrangian, 
we always use the proton
mass $M_p$ as scaling mass.
With this, we follow established conventions, as 
discussed above in conjunction
with the pion Langrangians.
We note that disregarding this point would generate noticable, 
but unmotivated CSB.

Finally, the Lagrangians for $\omega$ and $\sigma'$ are:
\begin{eqnarray}
{\cal L}_{\omega NN} &  = &  g_{\omega NN}
                           \bar{\psi}  \gamma_\mu \psi
                           \varphi_\omega^\mu \; ,
\\
{\cal L}_{\sigma' NN} &  = &  g_{\sigma' NN}
                           \bar{\psi}  \psi
                           \varphi_{\sigma'} \; .
\end{eqnarray}

Starting from these Lagrangians, irreducible diagrams up to fourth order
are evaluated using old-fashioned/time-ordered perturbation theory.
Some important diagrams (but not all) are shown in Figs.~1--4.
The sum of all irreducible diagrams included in the model is, by definition, 
the quasi-potential $V$.
Mathematically, this quasi-potential is the kernel
of the scattering equation.
For an uncoupled partial wave with angular momentum $J$,
this equation reads:
\begin{equation}
R_J(q',q) =  V_J(q',q) + {\cal P}\int^{\infty}_0 
\frac{dk\: k^2}{2E_q-2E_k}\: V_J(q',k)\: R_J(k,q)
\end{equation}
with $q$, $k$, and $q'$ the magnitude of the relative momenta
of the two interacting nucleons in the initial, intermediate,
and final state, respectively; 
$E_q=\sqrt{M^2+q^2}$
and
$E_k=\sqrt{M^2+k^2}$
 with
$M$ the correct mass of the nucleon involved in the
scattering process under consideration.
The principal value is denoted by $\cal P$ and $R$ is
commonly called the K-matrix.
By solving this equation,
the kernel/quasi-potential is iterated infinitely many times. 
This is equivalent to solving the Schroedinger equation.

From the on-shell $R$-matrix, phase shifts for uncoupled
partial waves are obtained through:
\begin{equation}
\tan\: \delta_J(T_{lab}) = -\frac{\pi}{2} q E_q\: R_J(q,q)
\end{equation}
where $q$ denotes the on-shell momentum in the center-of-mass
system of the two nucleons which is related to the laboratory 
kinetic energy by $T_{lab}=2q^2/M$.

Further details concerning the formalism can be found in
appendices A to C of Ref.~\cite{MHE87}.

\section{CSB due to nucleon mass difference}

It is the purpose of the present investigation to take the
nucleon mass splitting accurately into account,
which leads to CSB. Therefore, we use exact values for 
the proton mass $M_p$ and neutron mass $M_n$~\cite{PDG92}:
\begin{eqnarray}
M_p = 938.2723 \mbox{ MeV} , 
\\
M_n = 939.5656 \mbox{ MeV} .
\end{eqnarray}

We start with $pp$ scattering for which
the one-boson-exchange contribution is depicted in Fig.~1b
and $2\pi$-exchange contributions are shown in Fig.~2b, 3b, and 4b.
Note that in all of these diagrams, the proton carries the
exact proton mass $M_p$ and the neutron,
which occurs in some intermediate states, carries the exact
neutron mass $M_n$.
For the $\Delta$-isobars, that are excited in some intermediate states
in Figs.~3 and 4, the average mass $M_\Delta = 1232$ MeV is used.

For the $pp$ case, our model yields $-17.20$ fm for the singlet
scattering length and 2.88 fm for the corresponding effective
range, consistent with Eqs.~(1) and (2).

Switching now---step by step---from $pp$ to $nn$ scattering will change the
effective range parameters and the phase shifts, 
in violation of charge-symmetry. 

The differences that occur for the effective range parameters are given
in Table~I and II.
Note that the relationship between the CSB potential and the
corresponding change of the
scattering length, $\Delta a_{CSB}$, is highly non-linear. 
As discussed in Refs.~\cite{EM83,CM86}, when the scattering
length changes from $a_1$ to $a_2$ due to a CSB potential
$\Delta V=V_1 - V_2$, the relationship is
\begin{equation}
\frac{1}{a_2} - \frac{1}{a_1} = M_N
 \int_0^\infty \Delta V u_1 u_2 dr
\end {equation}
or
\begin{equation}
a_1 - a_2 = a_1 a_2 M_N
 \int_0^\infty \Delta V u_1 u_2 dr \; ,
\end {equation}
with $u_1$ and $u_2$ the zero-energy $^1S_0$ wave functions
normalized such that $u(r\rightarrow \infty ) \longrightarrow
(1-r/a)$. Thus, the perturbation expansion concerns the
invers scattering length.
As clearly evident from Eq.~(20),
the change of the scattering length 
depends on the ``starting value'' $a_1$
to which the effect is added.
In our calculations, CSB effects are generated step by step,
which implies that the starting value $a_1$ is different
for different CSB effects.
This distorts the relative size of
the scattering length differences.
To make the relative comparison meaningful,
we have rescaled our results for $\Delta a_{CSB}$
according to a prescription given by Ericson and Miller~\cite{EM83},
which goes as follows.
Assume the ``starting value'' for the scattering length is $a_1$
and a certain CSB effect brings it up to $a_2$. 
Then, the resulting scattering length difference $(a_1-a_2)$ is rescaled by
\begin{equation}
\Delta a = (a_1 - a_2) \frac{a_{pp} a_{nn}}{a_1 a_2}
\end{equation}
with $a_{pp}=-17.3$ fm and $a_{nn}=-18.8$ fm.
This will make $\Delta a$ independent of the choice
for $a_1$.
The numbers given in Table I and II for $\Delta a_{CSB}$
are all rescaled according to this prescription.

To state the effects of CSB on the $NN$ phase
shifts, we introduce for each  $LSJ$ state the CSB
phase shift difference $\Delta\delta^{LSJ}_{CSB}(T_{lab})$, defined by
\begin{equation}
\Delta\delta^{LSJ}_{CSB}(T_{lab}) \equiv
 \delta^{LSJ}_{nn}(T_{lab})
-
 \delta^{LSJ}_{pp}(T_{lab})
\end{equation}
where $\delta^{LSJ}_{nn}$ denotes the $nn$
and $\delta^{LSJ}_{pp}$ the $pp$
phase shifts (without electromagnetic effects), respectively.

The irreducible diagrams included 
in the quasi-potential/kernel can be subdivided into several groups. 
After discussing the effect from the kinetic energy, we will
describe each group of diagrams and the implications for CSB.
\begin{enumerate}
\item
{\bf Kinetic energy} (kin.\ en.).
The kinetic energy is smaller for the neutron because of its larger
mass. This reduces the magnitude of the
energy-denominator in Eq.~(15) 
for $nn$ scattering as compared to $pp$, thus, enhancing the
(attractive) integral term for $nn$. In addition, the factor
$E_q$ in Eq.~(16) is larger for the larger nucleon mass, which
results in an overall enhancement of the magnitude of the
$nn$ phase shifts. The combined effect yields larger $nn$ phase shifts 
as compared to $pp$ if the nuclear potential is attractive,
and vice versa if the nuclear potential is repulsive.
This can be understood more easily in the frame work of the radial
Schroedinger equation in which the effective potential
is $MV$. Thus, no matter if the nuclear potential $V$ is attractive
or repulsive, its effect on the phase shifts is always
enhanced for the larger nucleon mass $M$. 
This explains why in $^3P_1$ the CSB phase
shift splitting, Eq.~(22), comes out negative (repulsive
potential, negative phase shift), while it is positive in all
other partial waves listed in Table III (column `kin.en.')
where the potentials are attractive (positive phase shifts).
The magnitude
of the singlet scattering length increases by 0.25 fm (cf.\ Table I, 
column `kin.en.')
for $nn$ scattering as compared to $pp$. 
This is, of course, well known, and the effect on the
scattering length is usually quoted
to be 0.30 fm~\cite{Hen69}.
Our value is slightly smaller which can be attributed
to the use of relativistic kinetic energies in our model.

\item
{\bf One-boson-exchange} (OBE, Fig.~1) contributions
mediated by $\pi^{0}(135)$, $\rho^0(770)$,
 $\omega(782)$, $a_0/\delta(980)$, and
$\sigma'(550)$. 
In the Bonn model~\cite{MHE87},
the $\sigma'$ describes only the correlated $2\pi$ exchange in 
$\pi\pi-S$-wave (and not the uncorrelate $2\pi$ exchange
since the latter is calculated explicitly, cf.\ Figs.~2--4).
Charge-symmetry is broken by the fact that
for $pp$ scattering the proton mass is used in the Dirac spinors
representing the four external legs (Fig.~1b), 
while for $nn$ scattering the neutron mass
is applied (Fig.~1a).
The CSB effect from the OBE diagrams is extremely small
(cf.\ Table~I and III, column `OBE').

\item
{\bf $2\pi$-exchange with $NN$ intermediate states} ($2\pi NN$), Fig~2. 
Notice first that
only non-iterative diagrams are to be considered, since
the iterative ones 
are generated by the scattering equation (15) from
the OBE diagrams. In our calculations,
we include always all time-orderings (except those with
 anti-baryons in intermediate states);
to save space, we display, however, only a few characteristic
graphs in Fig.~2 (this is also true for all diagrams shown or 
discussed below).
Part (a) of Fig.~2 applies to $nn$ scattering, while part (b)
refers to $pp$ scattering.
Notice that when charged-pion exchange is involved, the intermediate-state
nucleon differs from that of the external legs. This is an important subtlety
that we account for accurately in our calculations; neglecting this effect
causes a systematic error in the order of 100\%.
Numerical results for this class of diagrams are given in Table~II
and IV, column `$2\pi NN$'.

\item
{\bf $2\pi$-exchange with N$\Delta$ intermediate states} 
($2\pi N\Delta$), Fig.~3.
This class of diagrams causes by far the largest CSB effect on
the scattering length (Table~II) as well as on the 
phase shifts (Table~IV).
Again, it is important in all of these diagrams to take the intermediate-state
nucleon mass correctly into account.

\item
{\bf $2\pi$-exchange with $\Delta\Delta$ intermediate states}
($2\pi\Delta\Delta$), Fig.~4. 
The effects are smaller than for $2\pi N\Delta$ because there are no 
nucleon intermediate states. Thus, the nucleon mass splitting affects
only the outer legs which typically results in a small effect.

\item
{\bf $\pi\rho$-exchange with $NN$ intermediate states} ($\pi\rho NN$). 
Graphically, 
the $\pi\rho NN$ diagrams can be obtained
by replacing in each diagram of Fig.~2, one pion by a
$\rho$-meson of the same charge state (because of this simple
analogy, we do not show the $\pi\rho$ diagrams explicitly here).
In our calculations, the CSB effects of the $\pi\rho$
diagrams with $NN$ intermediate states are taken into account
accurately. 
The effect is typically opposite to the one from $2\pi NN$ exchange.

\item
{\bf $\pi\rho$-exchange with N$\Delta$ intermediate states} 
($\pi\rho N\Delta$).
Concerning the $\pi\rho$ diagrams
with $\Delta$ intermediate
states a comment is in place. In the Bonn model~\cite{MHE87},
the crossed $\pi\rho$ diagrams with $\Delta$ intermediate
states are included in terms of an approximation.
It is assumed that they differ from the corresponding box diagrams
(i.~e., the diagrams on the left-hand side of Figs.~3 and 
the ones in the first row of Fig.~4a and 4b, 
but with one $\pi$ replaced by one $\rho$) 
only by the isospin factor. Thus, the $\pi\rho$
box diagrams with $\Delta$ intermediate states are multiplied by an
isospin factor that is equal to the sum of the isospin factors
for box and crossed box.
The $\pi\rho N\Delta$ effect is in general substantial and typically 
of the opposite sign as compared to $2\pi N\Delta$.

\item
{\bf $\pi\rho$-exchange with $\Delta\Delta$ intermediate states}
($\pi\rho\Delta\Delta$). 
The effects are very small.

\item
{\bf Further $3\pi$ and $4\pi$ contributions} ($\pi\sigma+\pi\omega$).
The Bonn potential also includes some $3\pi$-exchanges that can be
approximated in terms of $\pi\sigma$ diagrams and $4\pi$-exchanges
of  $\pi\omega$ type.
The sum of these contributions is small. 
These diagrams have $NN$ intermediate states (similar to Fig.~2,
but with one of the two exchanged pions replaced by an isospin-zero boson) 
and, thus, are of intermediate range.
Except for $^1S_0$, their effect is negligible.
\end{enumerate}

This finishes our detailed presentation of the relevant diagrams and their
CSB effects which are plotted in Figs.~5 and 6.
The total CSB splitting of the singlet scattering length amounts
to 1.58 fm (cf.\ last column of Table~I) which agrees well with
the empirical value $1.5\pm 0.5$ fm, Eq.~(5).
The sum of all CSB effects on phase shifts is given in 
the last column of Table~III and plotted by the solid curve in Fig.~5.
The largest total effect listed in Table~III is 1.8 deg in $^1S_0$ at 1 MeV. 
In the $S$-wave, the effect decreases with energy and is 0.15 deg at
300 MeV. In $P$ and $D$ waves the CSB effect on phase shifts increases
with energy and is typically in the order of 0.1 deg at 300 MeV.
We do not list our results for partial waves with $L \geq 3$, 
since the CSB effect becomes negligibly small for high $L$:
less than 0.02 deg at 300 MeV and 0.01 deg or less at 200 MeV
for $F$ and $G$ waves and even smaller for higher partial waves.

Since the pion is involved in almost all diagrams 
considered in this study,
the CSB effect depends on the $\pi NN$ coupling
constant. In the present calculations, we follow the Bonn 
model~\cite{MHE87}: we assume charge-independence of the
coupling constant and use 
$f^2_{\pi NN}/4\pi = 0.0795$ which, via Eq.~(10), translates into
$g^2_{\pi NN}/4\pi = 14.4$.
In recent years, there has been some controversy about the
precise value of the $\pi NN$ coupling constant.
Unfortunately, the problem is far from being settled.
Based upon $NN$ phase shift analysis,
the Nijmegen group~\cite{Sto93} advocates 
the `small' charge-independent
value
$g^2_\pi/4\pi = 13.5(1)$,
while a very recent determination
by the Uppsala group~\cite{Eri95} based upon high
precision $np$ charge-exchange data at 162 MeV seems to confirm
in the large `text book' value
$g^2_{\pi^\pm}/4\pi = 14.5(3)$.
Other recent determinations 
are in-between the two extremes:
The VPI group~\cite{Arn94} quotes
$g^2_\pi/4\pi = 13.77(15)$ from $\pi N$ and $NN$
analysis with no evidence for charge-dependence.
Bugg and Machleidt~\cite{BM95} obtain
$g^2_{\pi^\pm}/4\pi = 13.69(39)$ and
$g^2_{\pi^0}/4\pi = 13.94(24)$
from the analysis of $NN$ elastic data between
210 and 800 MeV.
Because of this large uncertainty in the $\pi NN$
coupling constant, it is of interest to know
how the CSB effects depends on this constant.
Naturally, the $2\pi$ contributions are proportional to
$g_\pi^4$~\cite{foot} and the $\pi\rho$ ones to $g_\pi^2$.
Since the two contributions carry (in general) opposite signs and 
vary in their relative magnitude from partial wave to partial wave,
there is no simple rule for how the total CSB effect
depends on $g_\pi$.
The value
$g^2_\pi/4\pi = 13.6$
is currently fashionable among the new generation
of high-precision $NN$ potentials~\cite{MSS96,Wir95,Sto94}.
For that reason, we have repeated our CSB calculations
using $g^2_\pi/4\pi = 13.6$ and find that the total $\Delta a_{CSB}$
is reduced by about 15\% as compared to the calculation using
$g^2_\pi/4\pi = 14.4$ (Table I). The phase shift differences
are reduced by roughly the same percentage in most 
partial waves. The exact numbers for
$g^2_\pi/4\pi = 13.6$
will be published elsewhere.

\section{Summary and conclusions}

Based upon the Bonn meson-exchange model for the $NN$ interaction,
we have calulated the CSB effects due to nucleon mass splitting
on the phase shifts of $NN$ scattering and the singlet 
effective range parameters.
We give results for partial waves up to $L=2$ and
laboratory energies below 300 MeV.

A remarkable finding is that the experimental
CSB difference in the singlet scattering length can be explained
from nucleon mass splitting alone.

Concerning phase shift differences, we find the largest in the $^1S_0$ state
where they are most noticable at
low energy; e.~g., at 1 MeV, the difference is 1.8 deg, indicating that the
{\it nn} nuclear force is more attractive than the {\it pp} one.
The $^1S_0$ phase shift difference decreases with increasing energy
and is about 0.15 deg at 300 MeV.

The CSB effect on the phase shifts of
higher partial waves is small;
in $P$ and $D$ waves,
typically in the order of 0.1 deg at 300 MeV and less
at lower energies.
This is substantially smaller than what is required phenomenologically
to solve the so-called $A_y$ puzzle in elastic nucleon-deuteron
scattering at low energies~\cite{WG92}.

The major part of the CSB effect comes from diagrams of
$2\pi$ exchange where those with $N\Delta$ intermediate states
make the largest contribution. 
We also study  the CSB effect from
irreducible diagrams that exchange a $\pi$ and $\rho$ meson.
To our knowledge, this class of diagrams has never before
been considered in any calculation of the CSB nuclear force.
We find that the $\pi\rho$ diagrams give rise to non-negligible
CSB contributions that are typically opposite to the $2\pi$
effects. In most partial waves, the $\pi\rho$ effect reduces 
the CSB from $2\pi$ exchange in the order of 50\%. 

Coon and Niskanen~\cite{CN96} have investigated the CSB effect on the
singlet scattering length from the diagrams of Figs.~2 and 3,
using a nonrelativistic model.
Their total result, $\Delta a_{CSB}=1.56$ fm
(applying the dTRS $NN$ potential~\cite{TRS75}
 and a cutoff mass of 1 GeV at the pion
vertices), agrees well with our total. However, there are large
differences in the details: from $2\pi NN$ and $2\pi N\Delta$, 
Coon and Niskanen obtain 1.28 fm and 0.24 fm, respectively;
while we get 0.37 fm and 1.85 fm, respectively.
Thus, the ratio of the two contributions is very different.
From Ref.~\cite{MHE87} it is known, that the $2\pi N\Delta$
contribution to the nuclear force is about four
times the one from $2\pi NN$.
It is reasonable to expect that the CSB effect scales roughly
with the size of the contribution that generates it.
This is true for our result, which is why we have confidence
in our findings.
In the Bonn model, a cutoff mass of 1.2 GeV is used at the
pion vertices, while Coon and Niskanen use 1 GeV. This may
explain why our overall contribution from $2\pi$ exchange
is larger. On the other hand, our model also includes the important
$\pi\rho$ diagrams (that are omitted in Ref.~\cite{CN96}),
which reduce the overall CSB effect.

From the diagrams displayed in Figs.~3 and 4 it is evident
that additional CSB could be created from $\Delta$-mass
splitting. Unfortunately, the charge-splitting of the
$\Delta(1232)$-baryon mass is not well known~\cite{PDG92}.
Since our present investigation is restricted to reliably
known baryon-mass splitting, we do not consider
any $\Delta$-mass splitting and use the average value for
the $\Delta$-mass (1232 MeV) throughout. It is, however, 
worthwhile to mention that our model includes everything 
needed for a systematic investigation of
CSB effcts caused by an assumed $\Delta$-mass splitting.
This may be an intersting topic for a future study.

Traditionally, it was believed that 
$\rho^0-\omega$ 
mixing explains essentially all CSB in the nuclear force.
However, recently some doubt has been cast on this paradigm.
Some researchers~\cite{GHT92,PW93,KTW93} found that 
$\rho^0-\omega$ exchange may have a substantial
$q^2$ dependence such as to cause this contribution to nearly vanish
in $NN$.
Our finding that the empirically known CSB in the 
nuclear force can be explained solely from nucleon mass splitting 
(leaving essentially no room for additional CSB contributions 
from $\rho^0-\omega$ mixing or other sources) fits well into this 
new scenario.
However, since the issue of the $q^2$ dependence of
$\rho^0-\omega$ exchange is by no means settled
(see Ref.~\cite{MO95} for discussion and more references), 
it is premature to draw any definite conclusions.

\vspace*{1cm}
This work was supported in part by the U.S. National Science Foundation
under Grant No.~PHY-9603097 
and by the Idaho State Board of
Education.

\pagebreak

\pagebreak

\onecolumn

\begin{table}
\caption{CSB differences of the $^1S_0$ effective range parameters
as explained in the text.
$2\pi$ denotes the sum of all $2\pi$-contributions and $\pi\rho$
the sum of all $\pi\rho$-contributions. TBE (non-iterative two-boson-exchange)
is the sum of $2\pi$, $\pi\rho$, and $(\pi\sigma+\pi\omega)$.
}
\begin{tabular}{lrrrrrrr}
               &kin.\ en.\ & OBE &$2\pi$ &
 $\pi\rho$ &
 $\pi\sigma+\pi\omega$& TBE & Total  \\
 \hline 
$\Delta a_{CSB}$ (fm) &
   0.246 & 0.013 & 2.888 & --1.537 & --0.034 & 1.316 & 1.575 \\
$\Delta r_{CSB}$ (fm) &
   0.004 & 0.001 & 0.055 & --0.031 & --0.001 & 0.023 & 0.027 
\end{tabular}
\end{table}

\vspace*{3cm}

\begin{table}
\caption{CSB differences of the $^1S_0$ effective range parameters
from $2\pi$ and $\pi\rho$ diagrams as explained in the text.
}
\begin{tabular}{lrrrrrrr}
               &
 $2\pi NN$&
 $2\pi N\Delta$&
 $2\pi \Delta\Delta$&
 $\pi\rho NN$ &
 $\pi\rho N\Delta$ &
 $\pi\rho \Delta\Delta$ &
 Sum  \\
 \hline 
$\Delta a_{CSB}$ (fm) &
   0.374 & 1.852 & 0.662 & --0.484 & --1.184 & 0.130 & 1.350 \\
$\Delta r_{CSB}$ (fm) &
   0.005 & 0.036 & 0.014 & --0.010 & --0.025 & 0.003 & 0.024
\end{tabular}
\end{table}

\begin{table}
\caption{CSB phase shift differences (in degrees)
as defined in Eq.~(22). Notation as in Table~I.
}
\begin{tabular}{rrrrrrrr}
T$_{lab}$ (MeV)&kin.\ en.\ & OBE &$2\pi$ &
 $\pi\rho$ &
 $\pi\sigma+\pi\omega$& TBE & Total  \\
 \hline 
\hline
\\ \multicolumn{8}{c}{$^1S_0$} \\
    1 & 0.287 & 0.015 & 3.417 &-1.856 &-0.041 & 1.520 & 1.822 \\
    5 & 0.162 & 0.010 & 1.850 &-1.007 &-0.022 & 0.810 & 0.982 \\
   10 & 0.104 & 0.006 & 1.409 &-0.773 &-0.018 & 0.618 & 0.727 \\
   25 & 0.066 & 0.004 & 0.995 &-0.585 &-0.014 & 0.396 & 0.466 \\
   50 & 0.053 & 0.003 & 0.778 &-0.460 &-0.011 & 0.291 & 0.347 \\
  100 & 0.036 & 0.004 & 0.585 &-0.378 &-0.008 & 0.199 & 0.239 \\
  150 & 0.019 & 0.006 & 0.567 &-0.387 &-0.006 & 0.174 & 0.198 \\
  200 & 0.015 & 0.021 & 0.565 &-0.407 &-0.004 & 0.154 & 0.190 \\
  300 & 0.005 & 0.029 & 0.562 &-0.446 &-0.001 & 0.116 & 0.149 \\
\hline
\\ \multicolumn{8}{c}{$^3P_0$} \\
    5 & 0.004 & 0.003 & 0.001 & 0.000 & 0.000 & 0.001 & 0.009 \\
   10 & 0.010 & 0.006 & 0.001 & 0.001 & 0.000 & 0.002 & 0.019 \\
   25 & 0.020 & 0.014 & 0.003 & 0.003 & 0.001 & 0.007 & 0.042 \\
   50 & 0.025 & 0.018 & 0.006 & 0.006 & 0.002 & 0.014 & 0.057 \\
  100 & 0.025 & 0.014 & 0.008 & 0.010 & 0.003 & 0.021 & 0.060 \\
  150 & 0.016 & 0.017 & 0.007 & 0.012 & 0.004 & 0.023 & 0.057 \\
  200 & 0.008 & 0.022 & 0.006 & 0.014 & 0.005 & 0.024 & 0.054 \\
  300 & 0.004 & 0.023 & 0.002 & 0.016 & 0.005 & 0.022 & 0.050 \\
\hline
\\ \multicolumn{8}{c}{$^3P_1$} \\
    5 &-0.002 &-0.001 & 0.002 &-0.001 & 0.000 & 0.001 &-0.002 \\
   10 &-0.004 &-0.001 & 0.006 &-0.002 & 0.000 & 0.004 &-0.002 \\
   25 &-0.011 & 0.001 & 0.017 &-0.006 & 0.000 & 0.011 & 0.000 \\
   50 &-0.017 & 0.002 & 0.044 &-0.019 & 0.000 & 0.025 & 0.010 \\
  100 &-0.025 & 0.008 & 0.092 &-0.046 & 0.000 & 0.045 & 0.028 \\
  150 &-0.033 & 0.016 & 0.139 &-0.081 & 0.000 & 0.058 & 0.041 \\
  200 &-0.041 & 0.023 & 0.185 &-0.112 & 0.001 & 0.074 & 0.056 \\
  300 &-0.059 & 0.033 & 0.278 &-0.195 & 0.001 & 0.084 & 0.058 \\
\hline
\\ \multicolumn{8}{c}{$^1D_2$} \\
   25 & 0.001 & 0.001 & 0.002 & 0.000 & 0.000 & 0.002 & 0.004 \\
   50 & 0.004 & 0.001 & 0.008 &-0.001 & 0.000 & 0.007 & 0.012 \\
  100 & 0.007 & 0.002 & 0.031 &-0.007 & 0.000 & 0.024 & 0.033 \\
  150 & 0.011 & 0.003 & 0.061 &-0.018 & 0.000 & 0.043 & 0.057 \\
  200 & 0.012 & 0.003 & 0.095 &-0.034 & 0.000 & 0.061 & 0.076 \\
  300 & 0.014 & 0.003 & 0.178 &-0.078 & 0.000 & 0.100 & 0.117 \\
\hline
\\ \multicolumn{8}{c}{$^3P_2$} \\
    5 & 0.001 & 0.000 & 0.002 & 0.000 & 0.000 & 0.002 & 0.003 \\
   10 & 0.002 & 0.001 & 0.006 &-0.001 & 0.000 & 0.005 & 0.007 \\
   25 & 0.005 & 0.002 & 0.023 &-0.006 & 0.000 & 0.018 & 0.025 \\
   50 & 0.014 & 0.002 & 0.054 &-0.015 & 0.000 & 0.040 & 0.056 \\
  100 & 0.023 & 0.001 & 0.114 &-0.036 & 0.001 & 0.079 & 0.102 \\
  150 & 0.026 & 0.001 & 0.154 &-0.055 & 0.002 & 0.101 & 0.128 \\
  200 & 0.025 & 0.000 & 0.177 &-0.068 & 0.003 & 0.112 & 0.137 \\
  300 & 0.023 & 0.000 & 0.237 &-0.095 & 0.003 & 0.144 & 0.167 \\
\end{tabular}
\end{table}

\begin{table}
\caption{CSB phase shift differences (in degrees)
as defined in Eq.~(22) from the various $2\pi$ and
$\pi\rho$-exchange contributions as defined in the text.}
\begin{tabular}{rrrrrrrr}
T$_{lab}$ (MeV)&
 $2\pi NN$&
 $2\pi N\Delta$&
 $2\pi \Delta\Delta$&
 $\pi\rho NN$ &
 $\pi\rho N\Delta$ &
 $\pi\rho \Delta\Delta$ &
 Sum  \\
 \hline 
\hline
\\ \multicolumn{8}{c}{$^1S_0$} \\
    1 & 0.424 & 2.184 & 0.808 &-0.592 &-1.418 & 0.154 & 1.561 \\
    5 & 0.224 & 1.190 & 0.436 &-0.317 &-0.776 & 0.086 & 0.843 \\
   10 & 0.164 & 0.909 & 0.336 &-0.242 &-0.597 & 0.067 & 0.636 \\
   25 & 0.099 & 0.648 & 0.248 &-0.182 &-0.452 & 0.049 & 0.410 \\
   50 & 0.059 & 0.514 & 0.204 &-0.138 &-0.366 & 0.044 & 0.318 \\
  100 & 0.012 & 0.406 & 0.168 &-0.105 &-0.317 & 0.044 & 0.207 \\
  150 &-0.005 & 0.392 & 0.180 &-0.106 &-0.331 & 0.049 & 0.180 \\
  200 &-0.020 & 0.395 & 0.190 &-0.108 &-0.360 & 0.061 & 0.158 \\
  300 &-0.065 & 0.405 & 0.223 &-0.113 &-0.413 & 0.080 & 0.117 \\
\hline
\\ \multicolumn{8}{c}{$^3P_0$} \\
    5 &-0.001 & 0.001 & 0.000 & 0.000 & 0.000 & 0.000 & 0.001 \\
   10 &-0.003 & 0.004 & 0.001 & 0.000 & 0.001 & 0.000 & 0.002 \\
   25 &-0.012 & 0.013 & 0.002 & 0.000 & 0.003 &-0.001 & 0.006 \\
   50 &-0.022 & 0.024 & 0.004 & 0.001 & 0.006 &-0.001 & 0.012 \\
  100 &-0.036 & 0.038 & 0.006 & 0.001 & 0.011 &-0.002 & 0.018 \\
  150 &-0.044 & 0.044 & 0.007 & 0.000 & 0.014 &-0.003 & 0.019 \\
  200 &-0.051 & 0.049 & 0.008 & 0.000 & 0.017 &-0.003 & 0.019 \\
  300 &-0.064 & 0.057 & 0.009 &-0.003 & 0.022 &-0.004 & 0.017 \\
\hline
\\ \multicolumn{8}{c}{$^3P_1$} \\
    5 & 0.001 & 0.001 & 0.000 & 0.000 & 0.000 & 0.000 & 0.001 \\
   10 & 0.002 & 0.003 & 0.000 &-0.001 &-0.001 & 0.000 & 0.004 \\
   25 & 0.006 & 0.011 & 0.001 &-0.002 &-0.004 & 0.000 & 0.011 \\
   50 & 0.013 & 0.029 & 0.002 &-0.006 &-0.013 & 0.000 & 0.025 \\
  100 & 0.024 & 0.063 & 0.005 &-0.014 &-0.032 & 0.000 & 0.046 \\
  150 & 0.032 & 0.100 & 0.007 &-0.024 &-0.056 &-0.001 & 0.059 \\
  200 & 0.038 & 0.139 & 0.009 &-0.036 &-0.074 &-0.002 & 0.073 \\
  300 & 0.050 & 0.217 & 0.011 &-0.051 &-0.143 &-0.001 & 0.083 \\
\hline
\\ \multicolumn{8}{c}{$^1D_2$} \\
   25 & 0.001 & 0.001 & 0.000 & 0.000 & 0.000 & 0.000 & 0.002 \\
   50 & 0.003 & 0.005 & 0.000 & 0.000 &-0.001 & 0.000 & 0.007 \\
  100 & 0.010 & 0.019 & 0.002 &-0.002 &-0.005 & 0.000 & 0.024 \\
  150 & 0.016 & 0.041 & 0.003 &-0.004 &-0.013 & 0.000 & 0.043 \\
  200 & 0.021 & 0.071 & 0.004 &-0.005 &-0.029 & 0.000 & 0.061 \\
  300 & 0.027 & 0.142 & 0.009 &-0.011 &-0.065 &-0.001 & 0.100 \\
\hline
\\ \multicolumn{8}{c}{$^3P_2$} \\
    5 & 0.000 & 0.001 & 0.000 & 0.000 & 0.000 & 0.000 & 0.002 \\
   10 & 0.001 & 0.004 & 0.001 & 0.000 &-0.001 & 0.000 & 0.005 \\
   25 & 0.003 & 0.015 & 0.006 &-0.002 &-0.003 &-0.001 & 0.018 \\
   50 & 0.005 & 0.035 & 0.014 &-0.005 &-0.007 &-0.003 & 0.039 \\
  100 & 0.006 & 0.075 & 0.033 &-0.013 &-0.016 &-0.007 & 0.078 \\
  150 & 0.005 & 0.102 & 0.047 &-0.019 &-0.025 &-0.011 & 0.099 \\
  200 & 0.003 & 0.120 & 0.054 &-0.022 &-0.032 &-0.014 & 0.109 \\
  300 &-0.001 & 0.155 & 0.083 &-0.031 &-0.044 &-0.021 & 0.142 \\
\end{tabular}
\end{table}

\pagebreak

\begin{figure}
\caption{One-boson-exchange (OBE) contributions to 
(a) $nn$ and (b) $pp$ scattering.}
\end{figure}

\begin{figure}
\caption{Irreducible 2$\pi$-exchange diagrams with $NN$ intermediate
states for (a) $nn$ and (b) $pp$ scattering.}
\end{figure}

\begin{figure}
\caption{2$\pi$-exchange contributions with $N\Delta$ intermediate
states to (a) $nn$ and (b) $pp$ scattering.}
\end{figure}

\begin{figure}
\caption{2$\pi$-exchange contributions with $\Delta\Delta$ intermediate
states to (a) $nn$ and (b) $pp$ scattering.}
\end{figure}

\begin{figure}
\caption{CSB phase shift differences $\Delta\delta^{LSJ}_{CSB}$ (in degrees)
as defined in Eq.~(22) for laboratory kinetic energies $T_{lab}$
below 300 MeV and partial waves with
$L\leq 2$.
The CSB effects due to the kinetic energy, OBE, the entire $2\pi$ model,
and $\pi\rho$ exchanges
are shown by the dotted, dash-triple-dot, dashed, and dash-dot
curves, respectively.
The solid curve is the sum of all CSB effects.
(See text for further explanations.)}
\end{figure}

\begin{figure}
\caption{Similar to Fig.~5, but here the individual contributions
from the 2$\pi$ and $\pi \rho$ exchange are shown.
The CSB effects due to the $2\pi NN$, $2\pi N\Delta$, 
$2\pi \Delta\Delta$, $\pi\rho NN$, and $(\pi\rho N\Delta +
\pi\rho\Delta\Delta)$ diagrams
are shown by the dashed, solid, dotted, dash-dot, and dash-triple-dot
curves, respectively.
(See text for further explanations.)}
\end{figure}


\begin{thebibliography}{99}
\bibitem{foot}
Note that in the Bonn model~\cite{MHE87} 
$f_{\pi N\Delta}$ depends on $f_{\pi NN}$ via the $SU(3)$ based
relationship
\begin{displaymath}
f^2_{\pi N\Delta} = \frac{72}{25} f^2_{\pi NN} \; .
\end{displaymath}
\end{thebibliography}
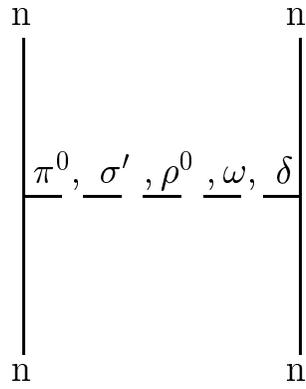
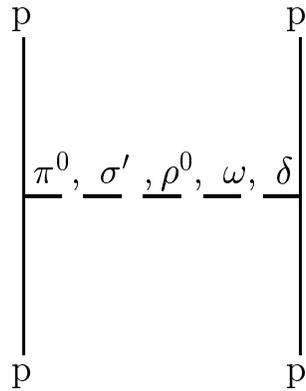
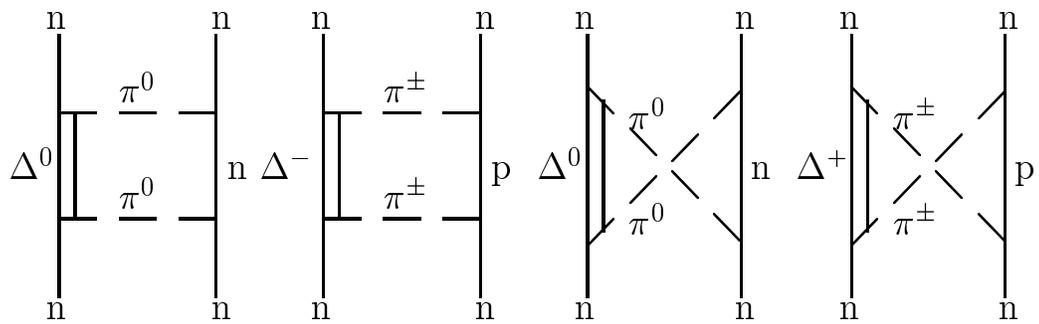
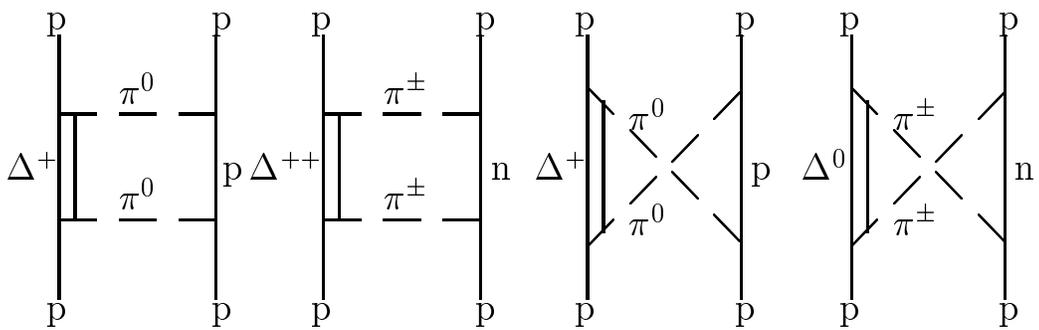
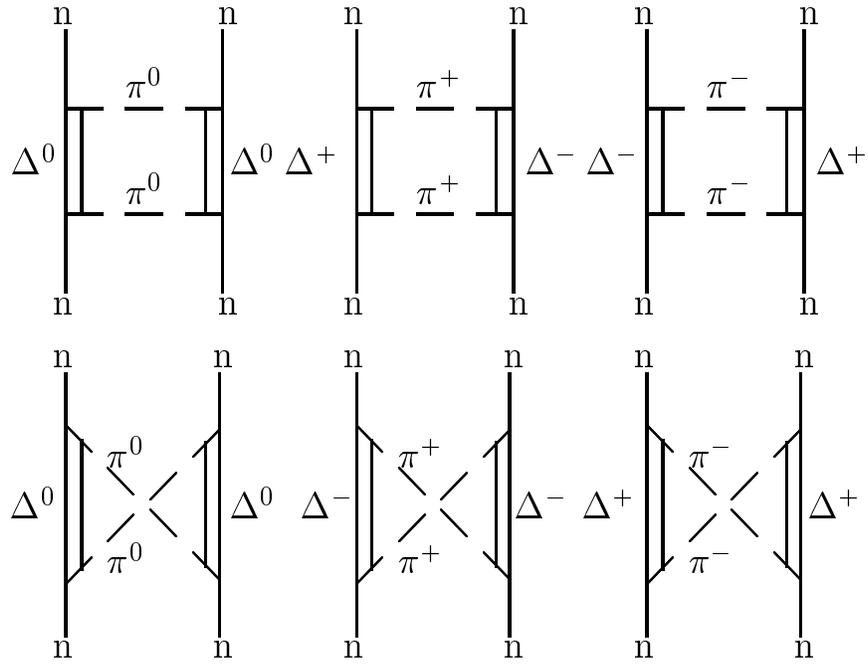
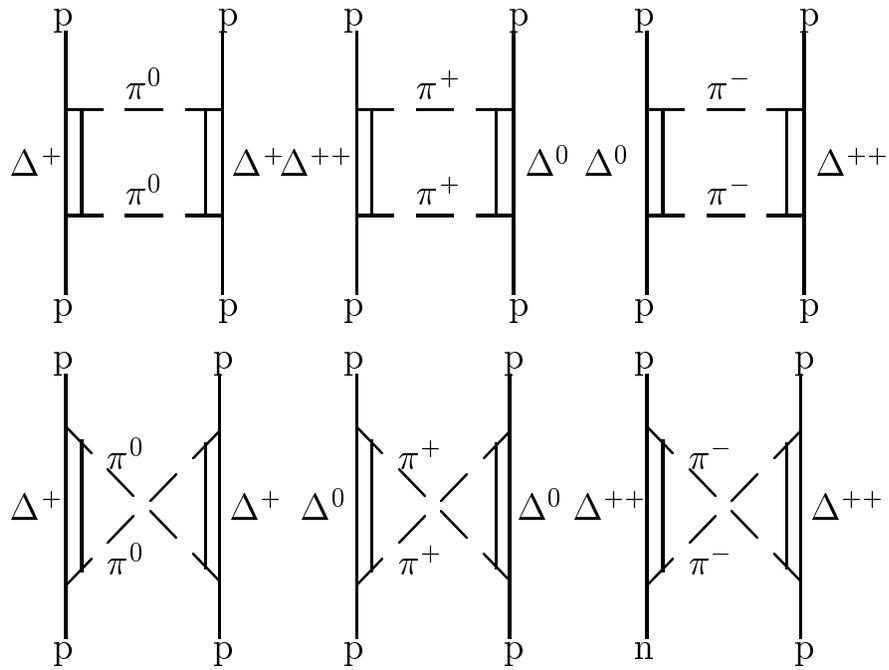
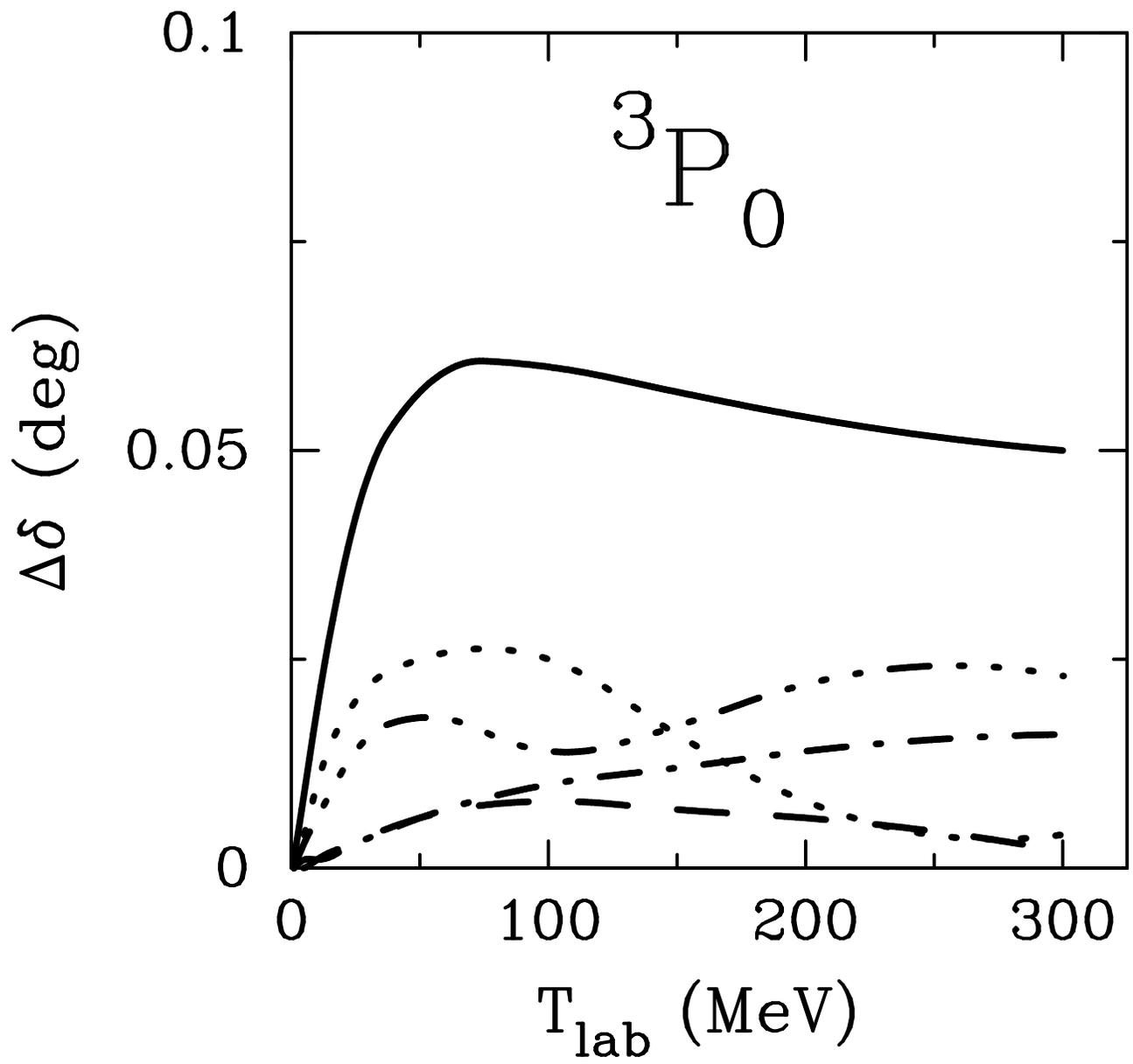
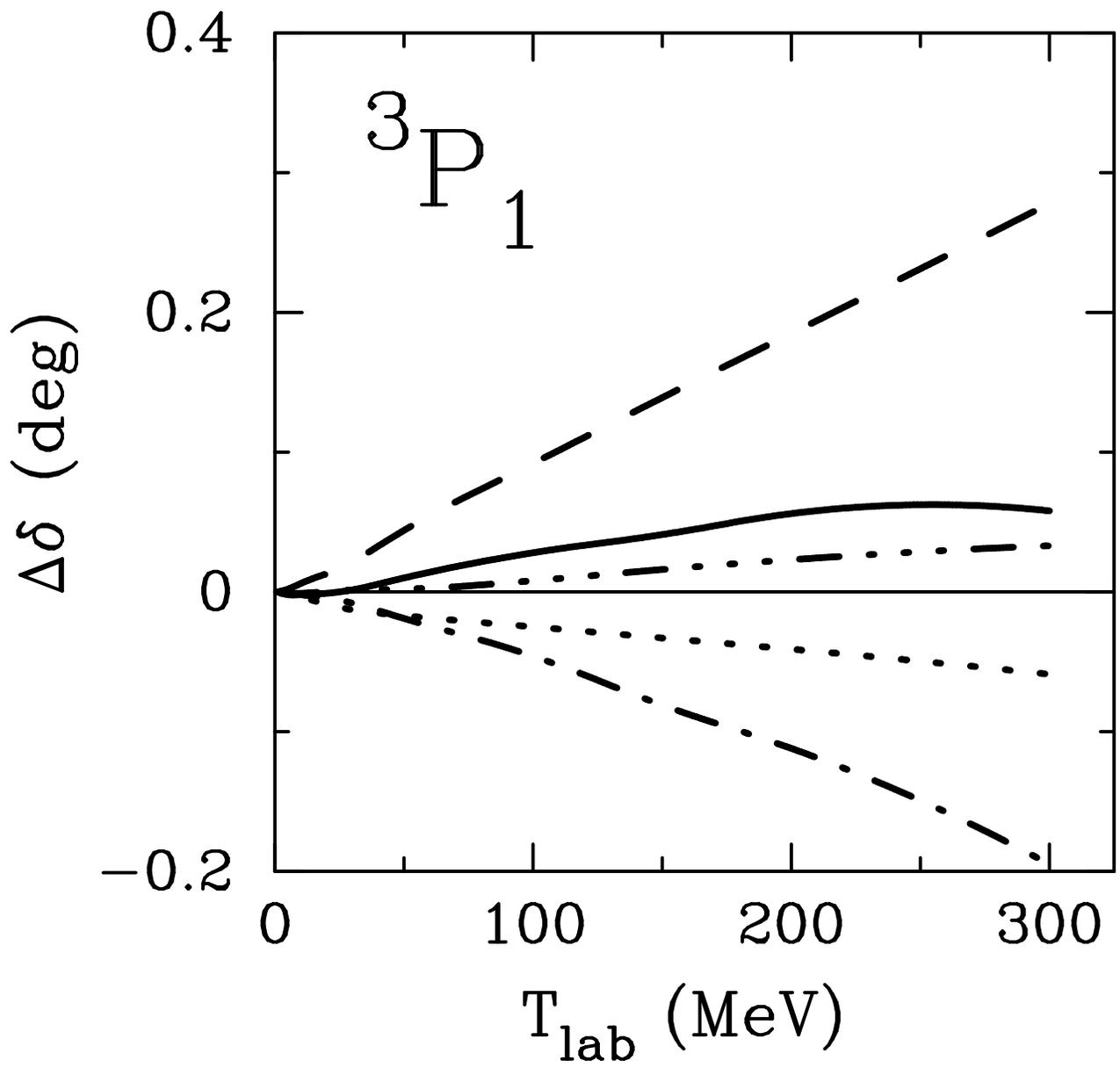
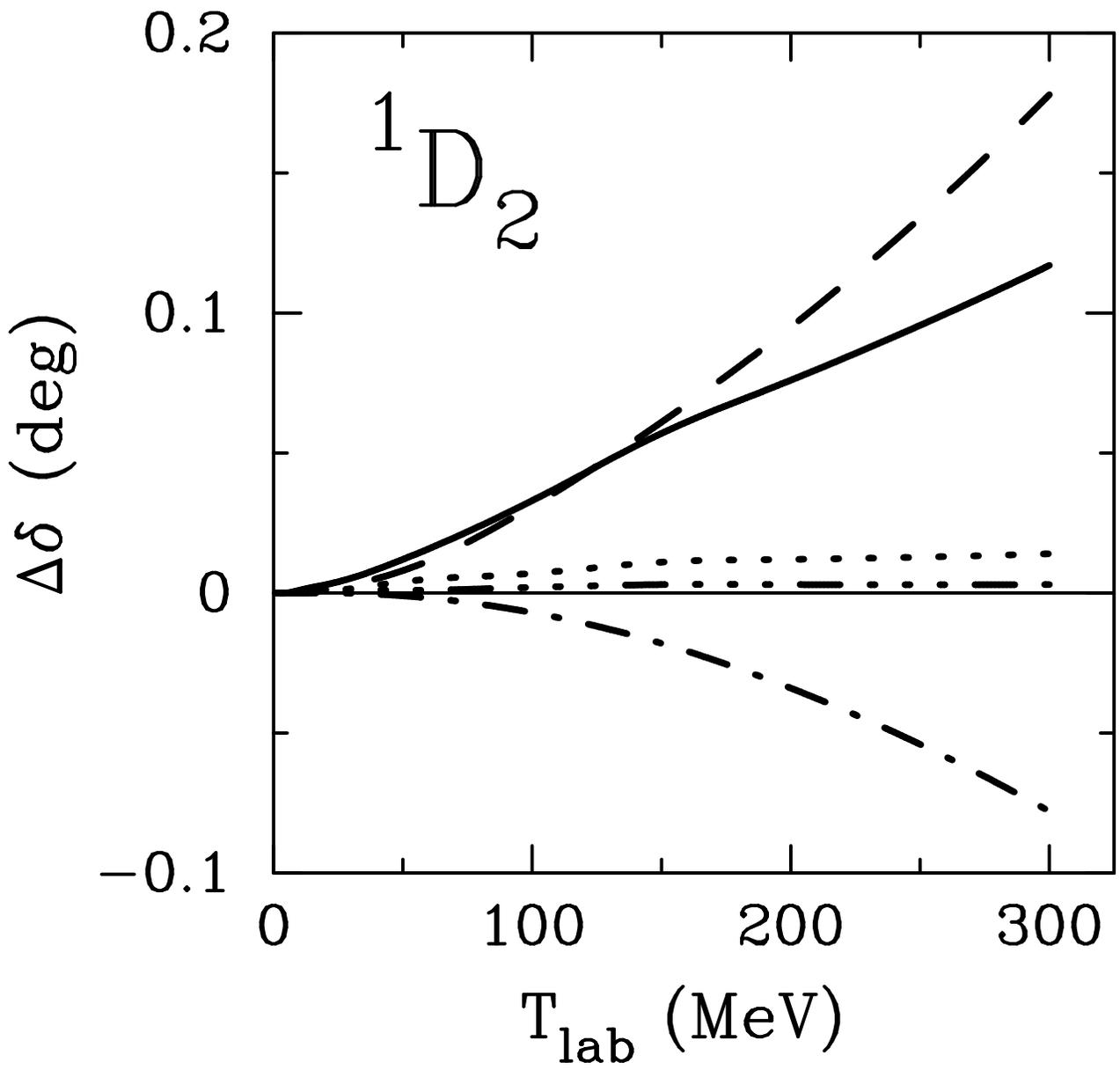
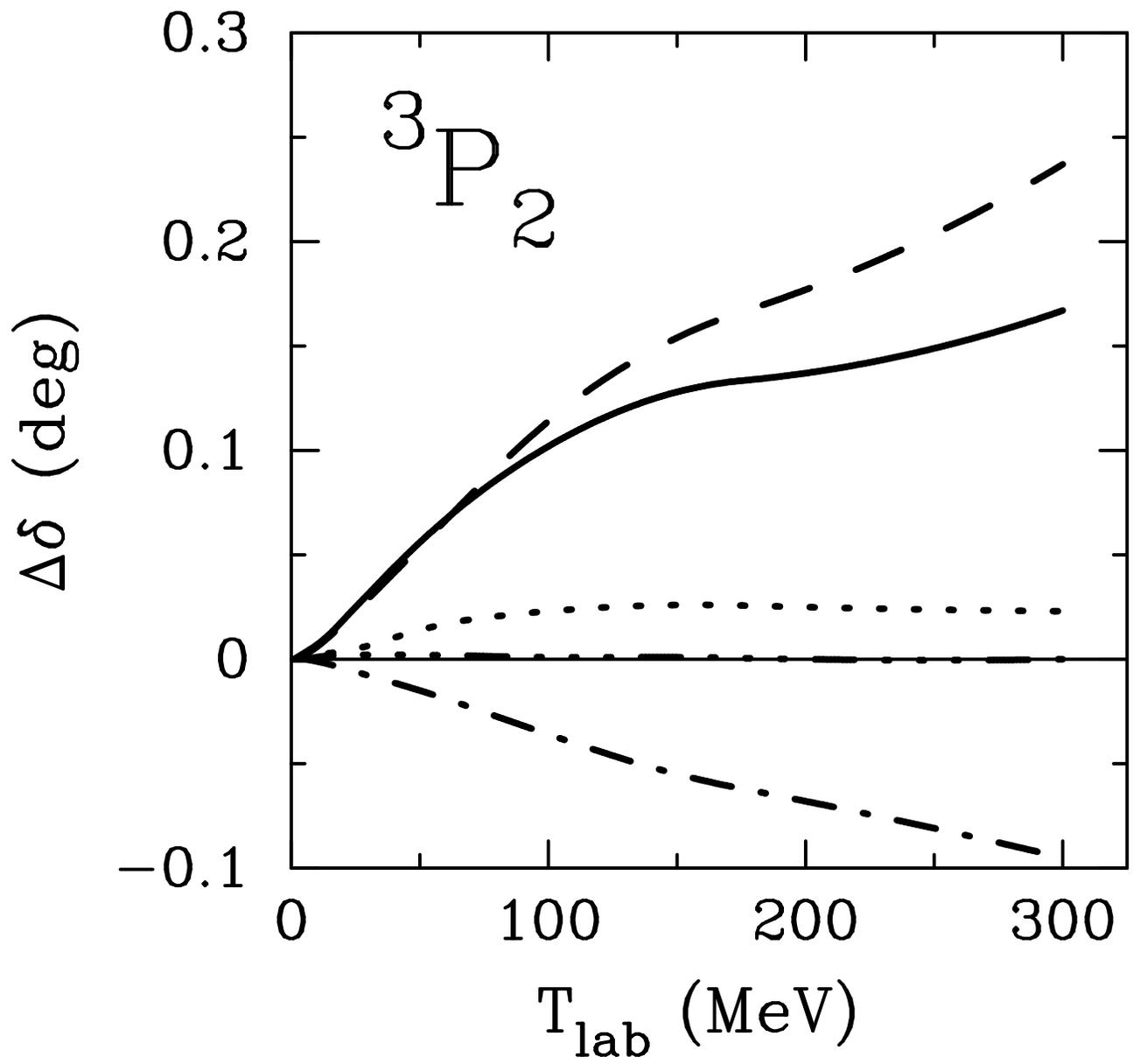
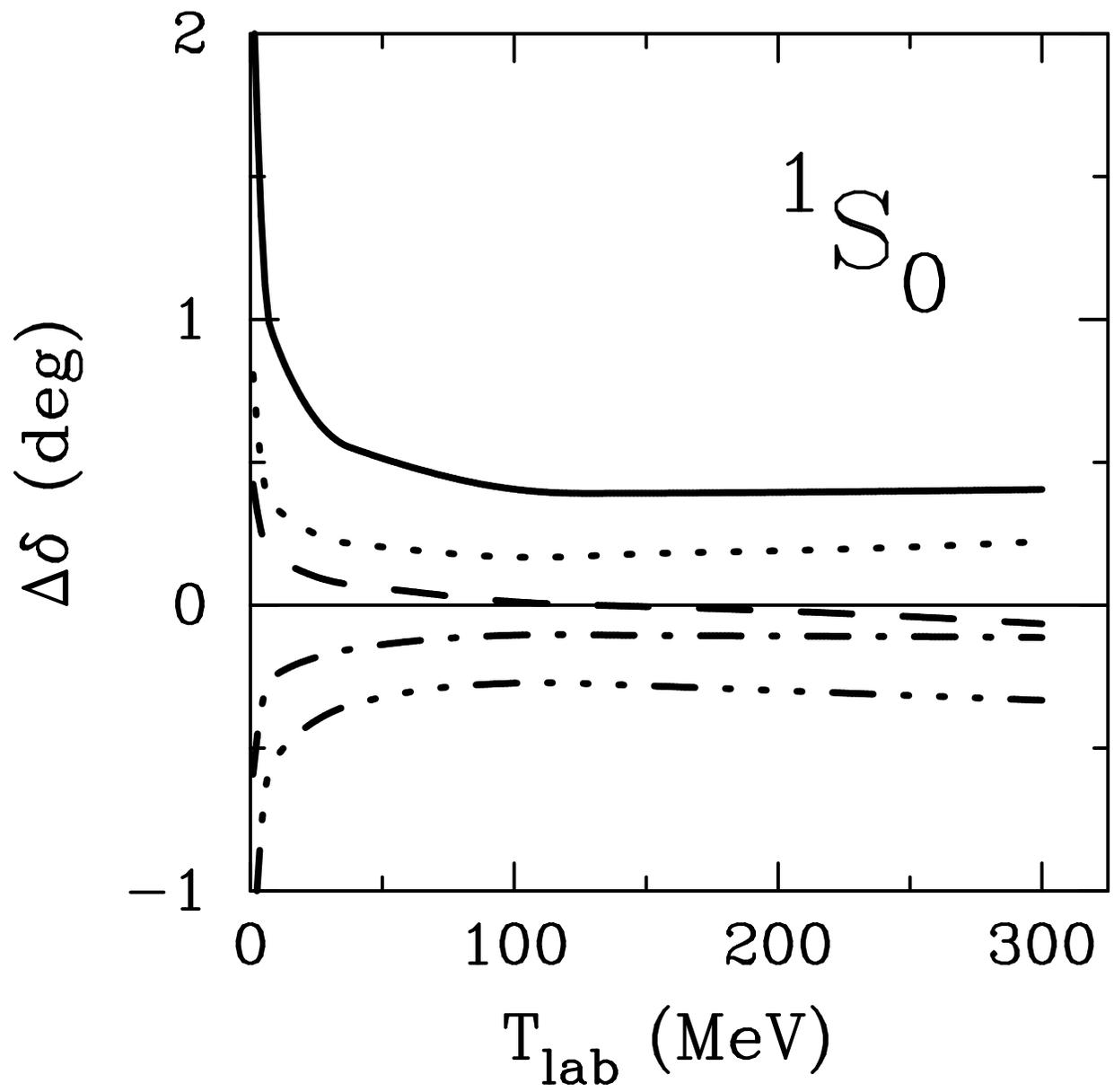
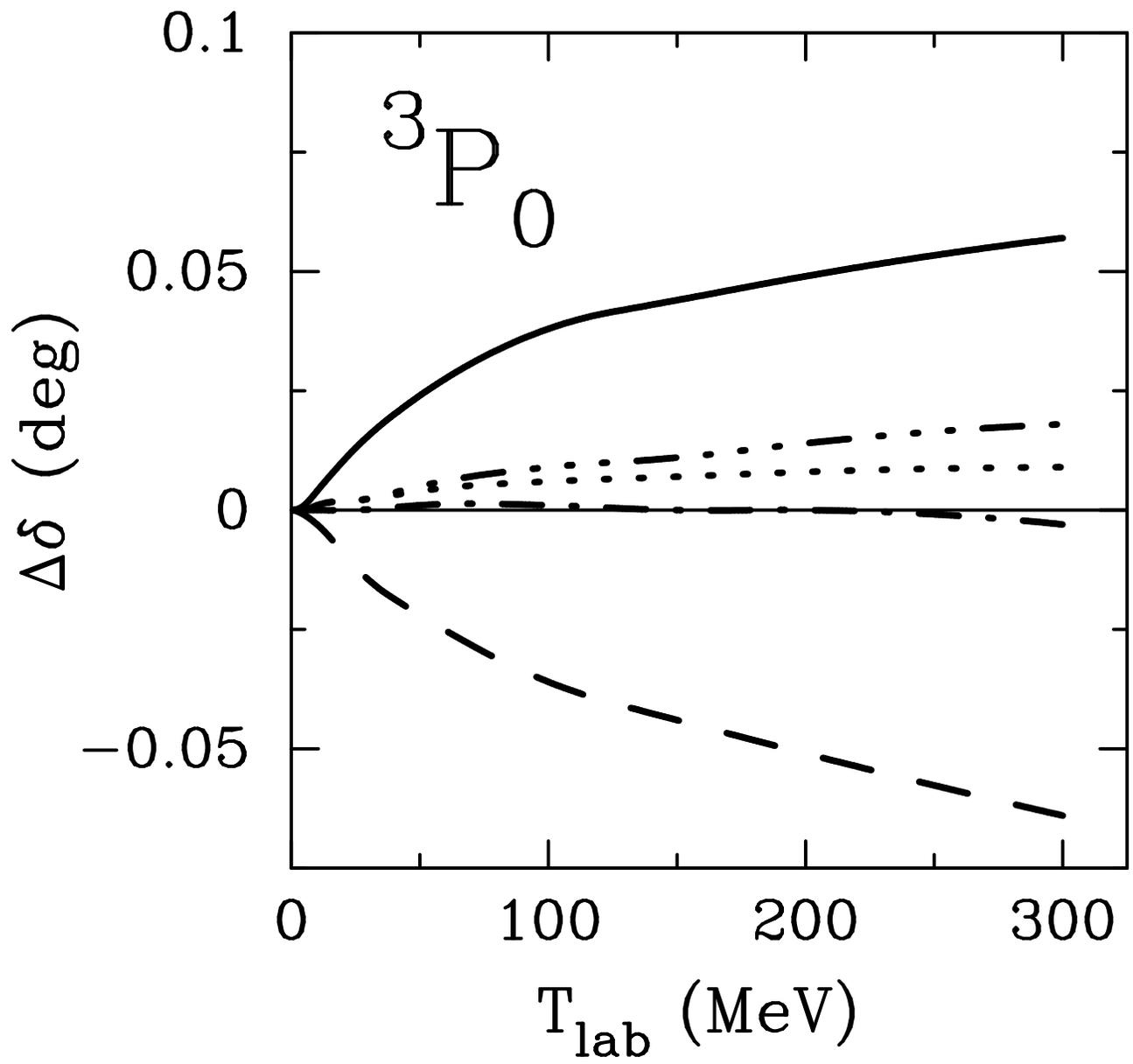
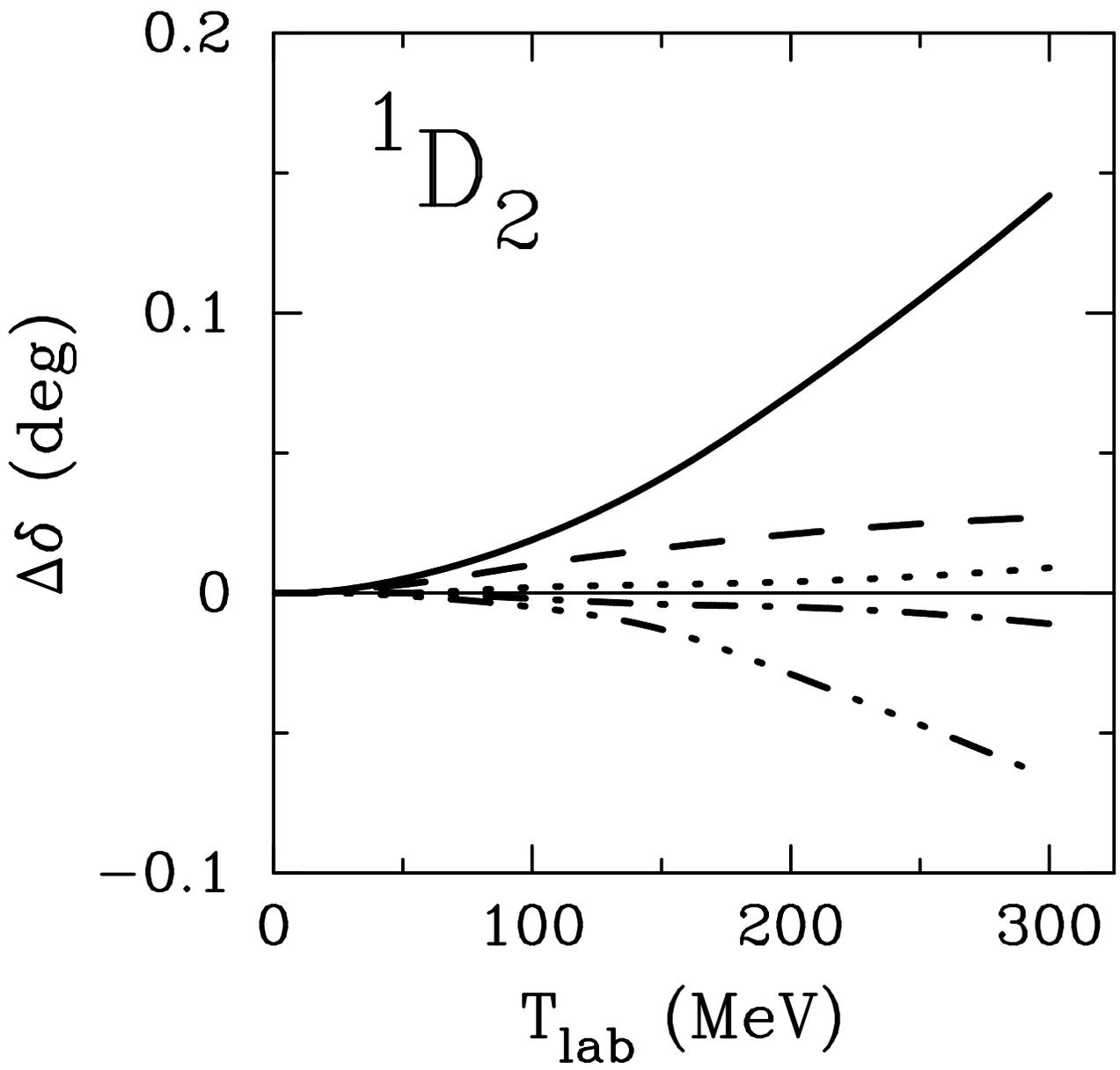
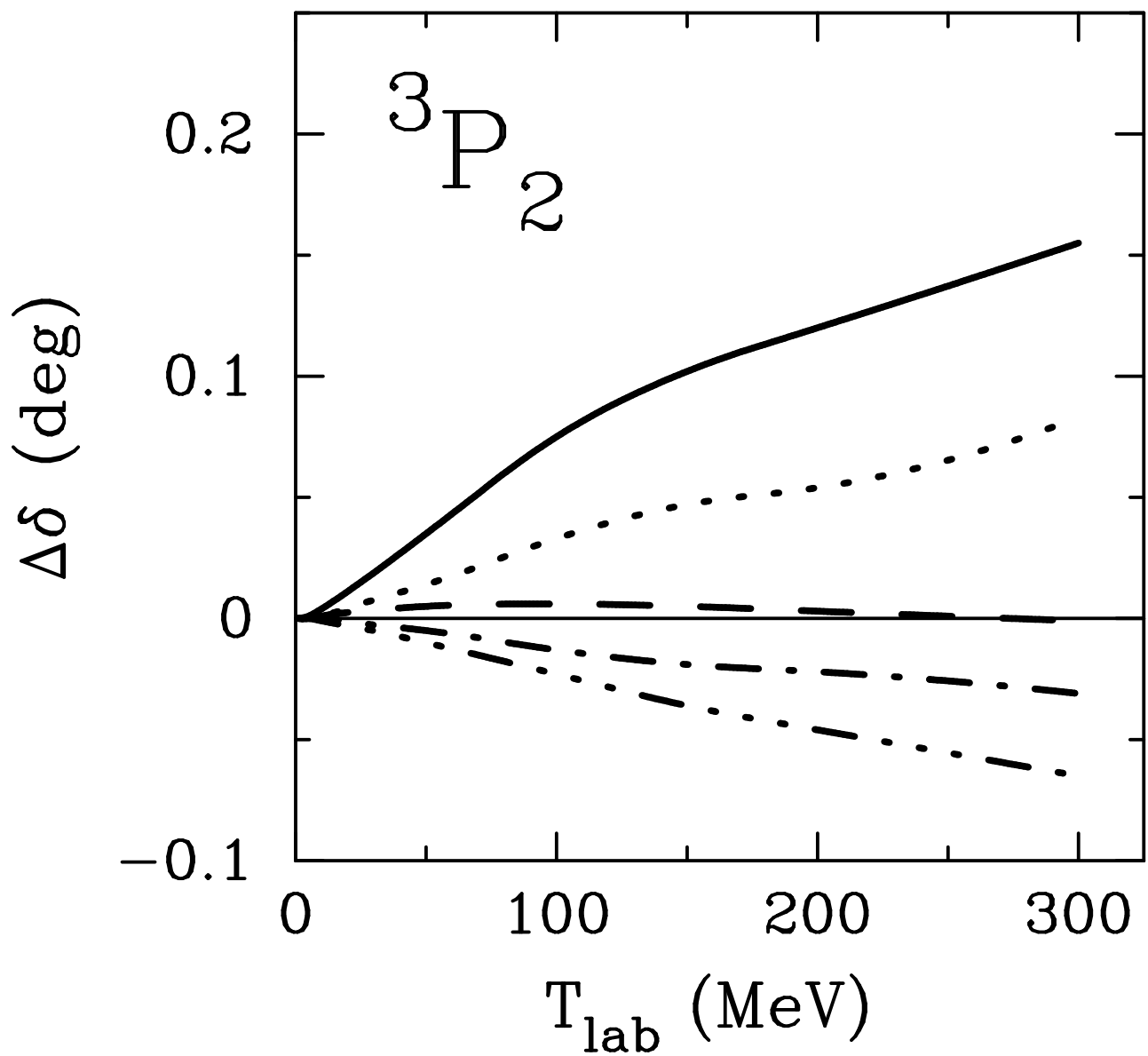
\end{document}